\documentclass[12pt]{article}
\usepackage{amsmath,amsfonts,amssymb}
\newcommand{\be}{\begin{equation}}
\newcommand{\ee}{\end{equation}}
\newcommand{\ba}{\begin{eqnarray}}
\newcommand{\ea}{\end{eqnarray}}

\newcommand{\la}[1]{\label{#1}}

\def\gl#1{(\ref{#1})}

\date{}
\begin{document}
\title{Spectral singularities for Non-Hermitian one-dimensional Hamiltonians:
puzzles with resolution of identity}
\author{A.A. Andrianov$^{1,2}$,\,F. Cannata$^{3}$,\,
A.V. Sokolov$^{1}$\\ {$^1$ \it Sankt-Petersburg State University, Russia}\\{$^2$ \it Institut de Ci\'encies del Cosmos UB, Barcelona, Spain}\\
{$^3$ \it INFN Bologna, Italy}} \maketitle \abstract{We examine
the completeness of biorthogonal sets of eigenfunctions for
non-Hermitian Hamiltonians possessing a spectral singularity. The
correct resolutions of identity are constructed for  delta like
and smooth potentials. Their form and the contribution of a
spectral singularity depend on the class of functions employed for
physical states. With this specification there is no obstruction to
completeness originating from a spectral singularity.}
\medskip

\begin{flushright}
{\sf Preprint ICCUB-10-013}
\end{flushright}
\medskip

\section{Introduction}
Lately complex Hamiltonians with real spectrum \cite{bender,acdi}
attract more and more attention to describe the phenomena in complex
crystals \cite{cryst}, in certain optical wave guides \cite{guides}
and in cosmology of dark energy \cite{cosm}. In the case of
discrete spectrum the peculiarities of non-Hermitian Hamiltonians
are related to the appearance of exceptional points due to
coalescence of some energy levels \cite{except}. For such systems
the complete biorthogonal set of eigen- and associated functions
normally exist and is sufficient to characterize their physics. On
the other hand, if a complex potential has bounded spatial
asymptotics, in the spectrum one can find not only continuum eigenvalues related to scattering
but also so called spectral singularities. The latter spectral
points lead to poles in the resolvent of the Hamiltonian in the
continuous part of the spectrum. This kind of spectral points are
known for a long time for radial problem of three dimensional
Schr\"odinger equation \cite{naim}-\cite{sams}.   Recently the states
corresponding to spectral singularities were discussed in
one-dimensional Quantum Mechanics on the entire real axis
\cite{mosta,ahmed} and for periodic complex potentials \cite{gusein,longhi} as
producing specific physical phenomena. However the observational
relevance of such states strongly depend on whether they appear as
independent building blocks in the complete set of biorthogonal
eigenstates.

In \cite{mosta,longhi} the serious doubts were raised concerning the very
existence of a complete resolution of identity in the case when
spectral singularities arise in the energy spectrum. In our work we
thoroughly examine this issue and arrive rather at the opposite
conclusion, namely, we build manifestly resolutions of identity for
typical complex potentials and point out how wave functions related to
spectral singularities are  incorporated in them. Meanwhile we have
found that the full contribution of eigenvectors of spectral
singularities is provided by different mechanisms and depends on a
class of test functions. In particular, for a narrower class of test
functions one can reduce the contribution of a spectral singularity
but at the expense of deletion of certain terms which are responsible
for reproducing of some test functions from a wider class.
The major part of this work including the Appendices is devoted to the rigorous justification of the completeness and the structure of resolutions of identities for different spaces of test functions.
We exemplify this reduction with an instructive example to elucidate
how different terms corresponding to the continuum and singular parts of the spectrum provide the identity.

The correct resolutions of identity are constructed not only for the
delta-like but also for some smooth potentials forming the
delta-like sequence.

\section{Resolution of identity for imaginary delta-like potential}

\hspace*{3ex} For continuous spectrum of the Hamiltonian \be
h=-\partial^2+z\delta(x),\qquad
\partial\equiv{d\over{dx}},\qquad iz\in \Bbb R\ee there are eigenfunctions
$$
\psi_+(x;k)={1\over\sqrt{2\pi}}\begin{cases}{{2k}\over{2k+iz}}\,e^{ikx},&x\geqslant0\\
e^{ikx}-{{iz}\over{2k+iz}}\,e^{-ikx},&x<0\end{cases}\equiv{1\over\sqrt{2\pi}}\Big(
e^{ikx}-{{iz}\over{2k+iz}}\,e^{ik|x|}\Big)$$
\be\equiv{1\over\sqrt{2\pi}}\Big(
e^{ikx}-{{iz}\over{2k+iz}}\,e^{-ikx}+{{2z}\over{2k+iz}}\,\theta(x)\sin
kx\Big),\la{p+}\ee $$
\psi_-(x;k)={1\over\sqrt{2\pi}}\begin{cases}(1+{{iz}\over{2k}})\,e^{-ikx}-{{iz}\over{2k}}\,
e^{ikx},&x\geqslant0\\
e^{-ikx},&x<0\end{cases}\equiv{1\over\sqrt{2\pi}}\Big(
(1+{{iz}\over{2k}})\,e^{-ikx}-{{iz}\over{2k}}\,e^{ik|x|}\Big)$$
\be\equiv{1\over\sqrt{2\pi}}\Big( e^{-ikx}+z\theta(x)\,{{\sin
kx}\over{k}}\Big),\la{p-}\ee \be h\psi_\pm=k^2\psi_\pm,\qquad
\theta(x)=\begin{cases}1,&x\geqslant0,\\
0,&x<0.\end{cases}\la{not4}\ee These functions satisfy to relations
\be\psi_-(-x;k)\equiv(1+{{iz}\over{2k}})\psi_+(x;k)\la{pp1}\ee and
\be
W[\psi_+(x;k),\psi_-(x;k)]\equiv\psi'_+(x;k)\psi_-(x;k)-\psi_+(x;k)\psi'_-(x;k)=
{{ik}\over\pi}.\ee
Let's notice that the standard "normalization" for scattering
is respected by $\psi_+$ where one can read off $T$ and $R$ ( transmission and reflection coefficients) but not by $\psi_-$ which in order to read off T and R would
require to be divided by $(1+iz/2k)$.

Green function for $h$ takes the form,
$$G(x,x';\lambda)={{\pi i}\over{\sqrt\lambda}}\,\psi_+(x_>;\sqrt\lambda)
\psi_-(x_<;\sqrt\lambda),\qquad{\rm{Im}}\,
\sqrt\lambda\geqslant0,$$ \be(h-\lambda)G=\delta(x-x'),\qquad
x_>=\max\{x,x'\},\quad x_<=\min\{x,x'\}.\la{gf}\ee

There is spectral singularity in the spectrum of $h$ for
$\lambda=-z^2/4$, which is the only pole of Green function \gl{gf}.
The corresponding eigenfunctions of $h$ take the form,
\be\psi_0(x)\equiv
e^{z|x|/2}=\sqrt{2\pi}\,\,\psi_-(x;-iz/2)=-\sqrt{2\pi}\lim_{k\to-iz/2}
[(1+{{2k}\over{iz}})\psi_+(x;k)],\la{p0}\ee
\be\psi_+(x;iz/2)\equiv{1\over\sqrt{2\pi}}\begin{cases}{1\over2}\,e^{-zx/2},&x\geqslant0\\
e^{-zx/2}-{1\over2}\,e^{zx/2},&x<0\end{cases}=
{1\over\sqrt{2\pi}}\,\psi_0(x)-{1\over2}\,\psi_-(x;iz/2)\ee and
\be\psi_-(x;iz/2)\equiv{1\over\sqrt{2\pi}}\begin{cases}2e^{zx/2}-
e^{-zx/2},&x\geqslant0\\
e^{zx/2},&x<0\end{cases}=\sqrt{2\over\pi}\,\psi_0(x)-2\psi_+(x;iz/2).\ee

The eigenfunctions $\psi_+(x;k)$ and $\psi_-(x;k)$ of $h$ satisfy
(see Appendix 1) the bi\-or\-tho\-go\-na\-li\-ty relations,
\be\int\limits_{-\infty}^{+\infty}[(1+{{2k}\over{iz}})\psi_+(x;k)]\psi_-(x;k')
\,dx=(1+{{2k}\over{iz}})\,\delta(k-k'),\la{dk}\ee where the
eigenfunction $\psi_0(x)$ is included due to \gl{p0}. 

The resolution of identity constructed from $\psi_+(x;k)$ and
$\psi_-(x;k)$ holds (see Appendix 2), \be\delta(x-x')=\int_{\cal
L}\psi_+(x;k)\psi_-(x';k)\,dk,\la{del}\ee where $\cal L$ is an
integration path in complex $k$ plane, obtained from the real axis by
its deformation near the point $k=-iz/2$ upwards\footnote{Alternatively one could
shift the denominators in \gl{p+} to $2k+iz+i0$. Then in \gl{del} one
can keep integration along real
axis.} and the direction of $\cal L$ is specified from
$-\infty$ to $+\infty$. This resolution of identity is valid for test functions
belonging to $C_{\Bbb R}\cap C^\infty_{(-\infty,0]}\cap
C^\infty_{[0,+\infty)}\cap L_2(\Bbb R;(1+|x|)^\gamma)$, $\gamma>-1$
as well as for some bounded and even slowly increasing test
functions (more details are presented in Appendix 2) and, in particular, for
eigenfunctions \gl{p+} and \gl{p-} of the Hamiltonian $h$.

One can rearrange the resolution of identity \gl{del} for any
$\varepsilon>0$ (see Appendix 2) to the form
$$\delta(x-x')=\Big(\int\limits_{-\infty}^{-iz/2-\varepsilon}
+\int\limits_{-iz/2+\varepsilon}^{+\infty}\Big)\,\psi_+(x;k)\psi_-(x';k)\,dk+$$
$$+{1\over\pi}\,e^{z(x-x')/2}{{\sin\varepsilon(x-x')}\over{x-x'}}+
{{iz}\over\pi}\,\theta(-x)\theta(x')\int\limits_{-iz/2-\varepsilon}^{-iz/2+\varepsilon}
{1\over k}\,\sin kx\,\sin kx'\,dk-$$
\be-{z\over4}\,\psi_0(x)\psi_0(x')\Big[1-{2\over\pi}
\int\limits_0^{\varepsilon(|x|+|x'|)}{{\sin t}\over
t}\,dt\Big],\qquad\varepsilon>0\la{ree15}\ee and, consequently, to
the form
$$\delta(x-x')={\lim_{\varepsilon\downarrow0}}'
\Big\{\Big(\int\limits_{-\infty}^{-iz/2-\varepsilon}
+\int\limits_{-iz/2+\varepsilon}^{+\infty}\Big)\,\psi_+(x;k)\psi_-(x';k)\,dk+$$
$$+{1\over\pi}\,e^{z(x-x')/2}{{\sin\varepsilon(x-x')}\over{x-x'}}+
{{iz}\over\pi}\,\theta(-x)\theta(x')\int\limits_{-iz/2-\varepsilon}^{-iz/2+\varepsilon}
{1\over k}\,\sin kx\,\sin kx'\,dk-$$
\be-{z\over4}\,\psi_0(x)\psi_0(x')\Big[1-{2\over\pi}
\int\limits_0^{\varepsilon(|x|+|x'|)}{{\sin t}\over
t}\,dt\Big]\Big\},\la{re16'}\ee where the prime $^\prime$ at the
limit symbol emphasizes that this limit is regarded as a limit in the space of
distributions.

We can reduce the resolution \gl{re16'} (see Appendix 2) for test
functions from $C_{\Bbb R}\cap C^\infty_{(-\infty,0]}\cap
C^\infty_{[0,+\infty)}\cap L_2(\Bbb R;(1+|x|)^\gamma)$, $\gamma>-1$
to the form
$$\delta(x-x')={\lim_{\varepsilon\downarrow0}}'\Big\{\Big(\int\limits_{-\infty}^{-iz/2-
\varepsilon}
+\int\limits_{-iz/2+\varepsilon}^{+\infty}\Big)\,\psi_+(x;k)\psi_-(x';k)\,dk-$$
$$
-{z\over4}\,\psi_0(x)\psi_0(x')\Big[1-{2\over\pi}\int\limits_0^{\varepsilon(|x|+|x'|)}{{\sin
t}\over t}\,dt\Big]\Big\}\equiv$$
\be{\lim_{\varepsilon\downarrow0}}'\Big\{\Big(\int\limits_{-\infty}^{-iz/2-\varepsilon}
+\int\limits_{-iz/2+\varepsilon}^{+\infty}\Big)\,\psi_+(x;k)\psi_-(x';k)\,dk-
{z\over{2\pi}}\,\psi_0(x)\psi_0(x')\int\limits_{\varepsilon(|x|+|x'|)}^{+\infty}{{\sin
t}\over t}\,dt\,\Big\}\la{re16}\ee and for test functions from
$C_{\Bbb R}\cap C^\infty_{(-\infty,0]}\cap
C^\infty_{[0,+\infty)}\cap L_2(\Bbb R;(1+|x|)^\gamma)$, $\gamma>1$
to the form $$\delta(x-x')=
{\rm{p.v.}}'\int\limits_{-\infty}^{+\infty}\psi_+(x;k)\psi_-(x';k)\,dk-
{z\over4}\,\psi_0(x)\psi_0(x')\equiv$$
\be{\lim_{\varepsilon\downarrow0}}'\Big(\int\limits_{-\infty}^{-iz/2-\varepsilon}
+\int\limits_{-iz/2+\varepsilon}^{+\infty}\Big)\,\psi_+(x;k)\psi_-(x';k)\,dk-
{z\over4}\,\psi_0(x)\psi_0(x').\la{re15}\ee

The resolution of identity \gl{re15} seems to have a more natural form than
\gl{re16} and especially \gl{re16'}, but formally, say, the right-hand part of
the resolutions \gl{re16} and \gl{re15} reproduces a half only of the function
$\psi_0(x)$ in view of \gl{dk} and of the following,
$$\lim_{\varepsilon\downarrow0}\int\limits_{-\infty}^{+\infty}\Big\{-{z\over4}\,
\psi_0(x)\psi_0(x')
\Big[1-{2\over\pi}\int\limits_0^{\varepsilon(|x|+|x'|)}{{\sin
t}\over t}\,dt\Big]\Big\}\psi_0(x)\,dx=$$
$$\lim_{\varepsilon\downarrow0}\Big\{-{z\over{\pi}}\,\psi_0(x')
\int\limits_0^{+\infty}dx\,e^{zx}
\int\limits_{\varepsilon(x+|x'|)}^{+\infty}dt\,{{\sin t}\over
t}\Big\}=$$
$$-{1\over{\pi}}\,\psi_0(x')\lim_{\varepsilon\downarrow0}\Big\{
-\int\limits_{\varepsilon|x'|}^{+\infty} {{\sin t}\over t}\,dt+
\int\limits_0^{+\infty}{{\sin\varepsilon(x+|x'|)}\over
{x+|x'|}}\,e^{zx}\,dx\Big\}=$$
$$\psi_0(x')\Big\{{1\over2}-{1\over\pi}\,e^{-z|x'|}
\lim_{\varepsilon\downarrow0}\int\limits_{\varepsilon|x'|}^{+\infty}
{{\sin\tau}\over\tau}\,e^{z\tau/\varepsilon}\,d\tau\Big\}=$$
$$\psi_0(x')\Big\{{1\over2}-{1\over\pi}\,e^{-z|x'|}
\lim_{\varepsilon\downarrow0}\Big[\int\limits_0^{+\infty}
{{\sin\tau}\over\tau}\,e^{z\tau/\varepsilon}\,d\tau-\int\limits_0^{\varepsilon|x'|}
{{\sin\tau}\over\tau}\,e^{z\tau/\varepsilon}\,d\tau\Big]\Big\}=$$
\be\psi_0(x')\Big\{{1\over2}+{\rm{sign}}\,(iz){{i}\over{2\pi}}\,e^{-z|x'|}
\lim_{\varepsilon\downarrow0}\ln{{|z|+\varepsilon}\over{|z|-\varepsilon}}\Big\}=
{1\over2}\,\psi_0(x'),\la{v19}\ee where the formulae 2.5.3.12,
2.5.13.20 and 2.5.13.21 from \cite{prud} are taken into account.
The missing one-half of the function $\psi_0(x)$ is provided by the second
and third terms of the right-hand part of the resolution of identity \gl{re16'}
due to the chain of equalities,
$$\lim_{\varepsilon\downarrow0}\!\int\limits_{-\infty}^{+\infty}\!\Big[{1\over\pi}
\,e^{z(x-x')/2}
{{\sin\varepsilon(x\!-\!x')}\over{x-x'}}\Big]\psi_0(x)\,dx\!=\!{1\over{2\pi}}
\lim_{\varepsilon\downarrow0}\lim_{A\to+\infty}\!\int\limits_{-A}^{A}\!
dx\,\psi_0(x)\!\int\limits_{-iz/2-\varepsilon}^{-iz/2+\varepsilon}\!dk\,e^{ik(x-x')}\!=$$
$${1\over{2\pi}}\lim_{\varepsilon\downarrow0}\lim_{A\to+\infty}
\int\limits_{-iz/2-\varepsilon}^{-iz/2+\varepsilon}dk\int\limits_{-A}^{A}
dx\,e^{ik(x-x')+z|x|/2}=$$
$${1\over{2\pi}}\lim_{\varepsilon\downarrow0}\lim_{A\to+\infty}
\int\limits_{-iz/2-\varepsilon}^{-iz/2+\varepsilon}e^{-ikx'}\Big[
{{e^{i(k-iz/2)A}-1}\over{i(k-iz/2)}}+{{1-e^{-i(k+iz/2)A}}\over{i(k+iz/2)}}\Big]\,dk=$$
$${1\over{2\pi}}\lim_{\varepsilon\downarrow0}\lim_{A\to+\infty}\Big[
e^{zx'/2}\int\limits_{-iz-\varepsilon}^{-iz+\varepsilon}e^{-i\tau
x'}{{e^{i\tau A}-1}\over{i\tau}}
\,d\tau+e^{-zx'/2}\int\limits_{-\varepsilon}^{\varepsilon}e^{-i\tau
x'}{{1-e^{-i\tau A}}\over{i\tau}}\,d\tau\Big]=$$
$${1\over{2\pi}}\lim_{\varepsilon\downarrow0}\Big\{
-e^{zx'/2}\int\limits_{-iz-\varepsilon}^{-iz+\varepsilon}e^{-i\tau
x'}{{d\tau}\over{i\tau}}
+e^{-zx'/2}\lim_{A\to+\infty}\Big[\int\limits_{-\varepsilon}^{\varepsilon}e^{-i\tau
x'}{{1-e^{-i\tau
A}}\over{i\tau}}\,d\tau\Big]\Big\}=$$
$${1\over{2\pi}}\,e^{-zx'/2}\lim_{\varepsilon\downarrow0}\lim_{A\to+\infty}
\Big[\int\limits_{-\varepsilon}^{\varepsilon}{{1-e^{-i\tau
A}}\over{i\tau}}\,d\tau+
\int\limits_{-\varepsilon}^{\varepsilon}{{e^{-i\tau
x'}-1}\over{i\tau}}(1-e^{-i\tau A})\,d\tau\Big]=$$
\be{1\over{2\pi}}\,e^{-zx'/2}\lim_{\varepsilon\downarrow0}
\Big[\lim_{A\to+\infty}\int\limits_{-\varepsilon}^{\varepsilon}{{\sin\tau
A}\over{\tau}}\,d\tau+
\int\limits_{-\varepsilon}^{\varepsilon}{{e^{-i\tau
x'}-1}\over{i\tau}}\,d\tau\Big] ={1\over2}\,e^{-zx'/2},\la{cep20}\ee
where Riemann theorem and the formula 2.5.3.12 from \cite{prud} are
used, as well as, the following relation is employed
\be\lim_{\varepsilon\downarrow0}\int\limits_{-\infty}^{+\infty}
\Big[{{iz}\over\pi}\,\theta(-x)\theta(x')\int\limits_{-iz/2-\varepsilon}^{-iz+\varepsilon}
\sin kx\,\sin kx'\,{{dk}\over k}\Big]\psi_0(x)\,dx=
{1\over2}\,\theta(x')(e^{zx'/2}-e^{-zx'/2}),\la{cep21}\ee which can
be derived in the same way as \gl{cep20} and, finally, due to the
identities
\be{1\over2}\,e^{-zx/2}+{1\over2}\,\theta(x)(e^{zx/2}-e^{-zx/2})
\equiv{1\over2}\,e^{z|x|/2}\equiv{1\over2}\,\psi_0(x).\ee Thus, the
resolution of identity \gl{re16'} maps the function
$\psi_0(x)$ entirely\footnote{It is interesting that contributions of the
second and third terms of the right-hand part of \gl{re16'} in the
resolution of identity are (see Remark 3 of Appendix 2) singular
discontinuous functionals whose supports consist of the only element
which is the infinity.} and there is no any paradox of  a "defectiveness" of reduced resolutions of identity because the function $\psi_0(x)$ does not belong to the reduced spaces of test functions $C_{\Bbb R}\cap C^\infty_{(-\infty,0]}\cap
C^\infty_{[0,+\infty)}\cap L_2(\Bbb R;(1+|x|)^\gamma)$, $\gamma>-1$ or even $C_{\Bbb R}\cap C^\infty_{(-\infty,0]}\cap
C^\infty_{[0,+\infty)}\cap L_2(\Bbb R;(1+|x|)^\gamma)$, $\gamma>1$.

\noindent{\bf Example 1.} In order to elucidate how the reduced
resolution  \gl{re15} provides identity let's apply it to the
function smoothing  $\psi_0(x)$, namely, \be \psi_0(x;\alpha) =
\exp\big(\frac12 (z-\alpha)|x|\big);\quad \alpha > 0. \ee In the
pointwise limit $\alpha \to 0$ this function tends to $\psi_0(x)$
but this limit is incompatible with the selected reductions of test
function spaces. The binorm of this function is well defined, \be
\int^{+\infty}_{-\infty} dx \Big(\psi_0(x;\alpha)\Big)^2 = -
\frac{2}{z - \alpha} , \ee and it might be taken as a possible
definition for the binorm of $\psi_0(x)$ in the limit $\alpha \to
0$. However as this limit is pointwise in  $x$ and does not preserve
test function spaces the question of what is a best
definition for the binorm of $\psi_0(x)$ remains open.

Now let's apply the two components of the resolution \gl{re15} to
$\psi_0(x;\alpha)$.  With a chain of lengthy but straightforward
calculations based on Eqs. \eqref{p+}, \eqref{p-}, \eqref{p0} one
can show that,
\[
\lim_{\varepsilon\downarrow0,A\rightarrow +\infty}
\Big(\int\limits_{-A}^{-iz/2-\varepsilon}
+\int\limits_{-iz/2+\varepsilon}^{A}\Big)\,dk\,\psi_-(x';k)\int\limits_{-\infty}^{+\infty}
dx\,\psi_+(x;k)\exp\big(\frac12(z-\alpha)|x|\big)-
\]
\[ -
{z\over4}\,\psi_0(x')\int\limits_{-\infty}^{+\infty}
\exp\big((z-\frac12\, \alpha)|x|\big)\, dx =
\]
\be = \left(\psi_0(x';\alpha) - \frac{\psi_0(x')}{2 -
(\alpha/z)}\right) +  \frac{\psi_0(x')}{2 - (\alpha/z)} =
\psi_0(x';\alpha). \ee For small $\alpha/|z| \ll 1$ the spectral
singularity contributes almost as much as the continuum part of the
spectrum but this contribution $\sim \psi_0(x)$ does not belong to
the reduced space of test functions and its role is solely to
compensate a similar piece from resolution of the continuum
spectrum. We notice also that when thinking about the operation
$\varepsilon\downarrow0, A\rightarrow +\infty$ and the limit $\alpha\downarrow0$ one finds different
results depending on their order as it follows from previous
discussion. In particular, one reproduces a half only of the function $\psi_0(x)$ in full accordance with \eqref{re15} if firstly the limit $\alpha\downarrow0$ is performed.

\smallskip

Let us now comment some technical subtleties in the above relations and note that the integral from the right-hand part of \gl{del}
is understood (see Appendix 2)on its order as follows: \be\int_{\cal
L}\psi_+(x;k)\psi_-(x';k)\,dk=\mathop{{\lim}'}\limits_{A\to+\infty}\int_{{\cal
L}(A)}\psi_+(x;k)\psi_-(x';k)\,dk,\la{int'23}\ee where ${\cal L}(A)$
is a path in complex $k$ plane, made of the segment $[-A,A]$ by
its deformation near the point $k=-iz/2$ upwards and the direction
of ${\cal L}(A)$ is specified from $-A$ to $A$. Since the integral
from the right-hand part of \gl{int'23} is a standard integral (not
a distribution), in view of \gl{ree15} the following relations take
place, $$\int_{\cal L}\psi_+(x;k)\psi_-(x';k)\,dk=$$
$$\mathop{{\lim}'}
\limits_{A\to+\infty}\Big[\lim_{\varepsilon\downarrow0}\Big(
\int\limits_{-A}^{-iz/2-\varepsilon}+\int\limits_{-iz/2+\varepsilon}^A\Big)
\psi_+(x;k)\psi_-(x';k)\,dk-{z\over4}\,\psi_0(x)\psi_0(x')\Big]=$$
$$\mathop{{\lim}'}\limits_{A\to+\infty}{\rm{p.v.}}
\int\limits_{-A}^A
\psi_+(x;k)\psi_-(x';k)\,dk-{z\over4}\,\psi_0(x)\psi_0(x')=$$
\be{\rm{p.v.}} \int\limits_{-\infty}^{+\infty}
\psi_+(x;k)\psi_-(x';k)\,dk-{z\over4}\,\psi_0(x)\psi_0(x') ,
\la{int'24}\ee where the limit for $\varepsilon\downarrow0$ (and
consequently "p.v.") is regarded as pointwise one (not as a limit in
a function space). The latter equality in \gl{int'24} is considered
as a definition. Thus, the resolution of identity,
\be\delta(x-x')={\rm{p.v.}} \int\limits_{-\infty}^{+\infty}
\psi_+(x;k)\psi_-(x';k)\,dk-{z\over4}\,\psi_0(x)\psi_0(x'),\la{rez26}\ee
holds (cf. with \gl{re15}) and moreover this resolution is
equivalent to \gl{del}, {\it i.e.} it is valid for all test
functions for which \gl{del} is valid.

The resolution of identity \gl{re16} contains both the
eigenfunctions $\psi_+(x;k)$ and $\psi_-(x';k)$ for positive $k$
which describe scattering, and the eigenfunctions $\psi_+(x;k)$ and
$\psi_-(x';k)$ for negative $k$, which are linear combinations of
scattering state vectors. It follows from the identities:
$$\psi_+(x;k)=-{{iz}\over{2k+iz}}\,\psi_+(x;-k)+{{4k^2}\over{4k^2+z^2}}\,
\psi_-(x;-k),\qquad x\in\Bbb R,\quad k\in\Bbb C,$$
\be\psi_-(x;k)=\psi_+(x;-k)-{{iz}\over{2k-iz}}\,\psi_-(x;-k),\qquad
x\in\Bbb R,\quad k\in\Bbb C.\la{soot18}\ee
We remind that the standard "normalization" for scattering
is respected by $\psi_+$ where one can read off $T$ and $R$ ( transmission and reflection coefficients) but not by $\psi_-$ which in order to read off T and R would
require to be divided by $(1+iz/2k)$.

With the help of
\gl{soot18} one can rearrange the resolution of identity \gl{re16} for
test functions from $C_{\Bbb R}\cap C^\infty_{(-\infty,0]}\cap
C^\infty_{[0,+\infty)}\cap L_2(\Bbb R;(1+|x|)^\gamma)$, $\gamma>-1$
(see Appendix 2) to the form
$$\delta(x-x')={\lim_{\varepsilon\downarrow0}}'\Big\{\Big(\int\limits_0^{|z|/2-\varepsilon}
+\int\limits_{|z|/2+\varepsilon}^{+\infty}\Big)\Big[\psi_+(x;k)\psi_+(x';-k)+
{{4k^2}\over{4k^2+z^2}}\,\psi_-(x;k)\psi_-(x';-k)\Big]\,dk-$$
\be-{z\over4}\,\psi_0(x)\psi_0(x')\Big[1-{2\over\pi}\int\limits_0^{\varepsilon
(|x|+|x'|)}{{\sin t}\over t}\,dt\Big]\Big\},\la{re19}\ee where the
eigenfunctions $\psi_+(x,k)$ and $\psi_-(x;k)$ correspond to scattering states,
or to the symmetric form
$$\delta(x-x')={\lim_{\varepsilon\downarrow0}}'\Big\{\Big(\int\limits_0^{|z|/2-\varepsilon}
\!\!\!+\!\!\!\int\limits_{|z|/2+\varepsilon}^{+\infty}\Big)\Big[{{iz}\over{2k-iz}}\,
\psi_+(x;k)\psi_+(x';k)+
{{4k^2}\over{4k^2+z^2}}\,\psi_+(x;k)\psi_-(x';k)+$$
$$+{{4k^2}\over{4k^2+z^2}}\,\psi_-(x;k)\psi_+(x';k)+{{4ik^2z}
\over{(4k^2+z^2)(2k+iz)}}\,\psi_-(x;k)\psi_-(x';k)\Big]\,dk-$$
\be{-z\over4}\,\psi_0(x)\psi_0(x')
\Big[1-{2\over\pi}\int\limits_0^{\varepsilon (|x|+|x'|)}{{\sin
t}\over t}\,dt\Big]\Big\},\la{re21}\ee where all eigenfunctions in
the integral over $k$ describe scattering.

Finally let us remark that resolutions of identity equivalent
to \gl{del} -- \gl{re16'} can be obtained from \gl{re19} --
\gl{re21} by supplementing them with the two following terms:
\be{2\over\pi}\,\cos{{iz(x-x')}\over2}\,{{\sin\varepsilon(x-x')}\over{x-x'}}-
{{iz}\over{4\pi}}\int\limits_{iz/2-\varepsilon}^{iz+\varepsilon}
{{e^{ik(|x|+|x'|)}}\over{k+iz/2}}\,dk.\ee This fact can be easily
checked with the help of the relation \gl{pred39} from Appendix 2.
As well resolutions of identity of the type \gl{rez26} can be
produced from \gl{re19} -- \gl{re21} by the replacement
$${\lim_{\varepsilon\downarrow0}}'\Big(\int\limits_0^{|z|/2-\varepsilon}
+\int\limits_{|z|/2+\varepsilon}^{+\infty}\Big)\to
{\rm{p.v.}}\int\limits_0^{+\infty}$$ and by neglecting the integral
$\int_0^{\varepsilon(|x|+|x'|)}\sin t\,dt/t$.

\section{Smooth potentials with spectral singularity}

In order to use the technique of Supersymmetric Quantum Mechanics let us consider the shifted Hamiltonian $h^+=h+z^2/4$. With the help of the
standard construction of linear SUSY one can transform $h^+$, using
the function $\psi_0(x)$ as a transformation function, into the
Hamiltonian \be
h^-=q^-q^+=-\partial^2+{z^2\over4}-z\delta(x)=(h^+)^\dag,\quad
q^\pm=\mp\partial-\chi(x),\quad
\chi(x)={{\psi'_0(x)}\over{\psi_0(x)}}={z\over2}\,{\rm{sign}}\,x.\la{sp22}\ee
It is easy to see that this SUSY construction is a limiting
case of  the linear SUSY construction with the smooth superpotential
$\chi_\alpha(x)=(z/2)\,{\rm{tanh}}\,\alpha x$ for
${\rm{Re}}\,\alpha\to+\infty$. The main elements of these constructions
are presented in the following table.\footnote{The asymptotics in
the second column of the table are valid in the case
${\rm{Re}}\,\alpha>0$ only.}

\

\

\hskip-7mm {\LARGE\begin{tabular}{| c | c | c |} \hline
&{\normalsize$\alpha\in\Bbb C,\quad{\rm{Re}}\,\alpha\ne0$}&
{\normalsize$\alpha=+\infty$}\\
\hline{\normalsize$q^\pm_\alpha=\mp\partial-\chi_\alpha(x)$}&
{\normalsize$\chi_\alpha(x)={z\over2}\,{\rm{tanh}}\, \alpha
x$}&{\normalsize
$\chi_{\infty}(x)\!\equiv\!\chi(x)\!=\!{z\over2}\,{\rm{sign}}\, x$}
\\\hline{\normalsize $q^\pm_\alpha\varphi^\pm_\alpha=0$,}&
{\normalsize$\varphi^-_\alpha(x)=[2\,{\rm{cosh}}\,\alpha
x]^{z/{(2\alpha)}}=$
}&{\normalsize$\varphi^-_{\infty}(x)\!\equiv\!\psi_0(x)\!=\!e^{z|x|/2}$,}\\
{\normalsize$\varphi^\pm_\alpha(x)=e^{\mp\int\chi_\alpha(x)\,dx}$}
&{\normalsize$e^{z|x|/2}[1+o(1)]$, $x\to\pm\infty$,}&\\
&{\normalsize$\varphi^+_\alpha(x)=[2\,{\rm{cosh}}\,\alpha
x]^{-{z/(2\alpha)}}=$
}&{\normalsize$\varphi^+_{\infty}(x)=e^{-z|x|/2}$}\\&{\normalsize$e^{-z|x|/2}[1+o(1)]$,
$x\to\pm\infty$}&\\\hline \!{\normalsize
$h^\pm_\alpha\!=\!q^\pm_\alpha
q^\mp_\alpha\!=\!-\partial^2\!+\!V^{\pm}_\alpha(x)$,}\!&
{\normalsize$V^\pm_\alpha(x)={z^2\over4}-{{z/2(z/2\mp\alpha)}\over{{\rm{cosh}}^2\alpha
x}}$}&{\normalsize$V^\pm_{\infty}(x)={z^2\over4}\pm z\delta(x)$}\\
{\normalsize$V^\pm_\alpha(x)=\chi^2_\alpha(x)\pm\chi'_\alpha(x)$}&&\\\hline
\end{tabular}}

\

\

In both cases $\alpha= +\infty$ and $\alpha\in\Bbb C$,
${\rm{Re}}\,\alpha\ne0$ the function $\varphi_\alpha^\mp(x)$ is  an
eigenfunction of the Hamiltonian $h^\pm_\alpha=q^\pm_\alpha
q^\mp_\alpha$ for the eigenvalue $E=0$ corresponding to the spectral
singularity (see the table) in the spectrum of $h^\pm_\alpha$.

An eigenfunction of the Hamiltonian \be
h_\alpha\equiv-\partial^2-{{z/2(z/2-\alpha)}\over{{\cosh}^2\alpha
x}}=h^+_\alpha-{z^2\over4}\la{gam'}\ee for an eigenvalue $k^2$
satisfies the differential equation
\be-\psi''-{{z/2(z/2-\alpha)}\over{{\cosh}^2\alpha
x}}\,\psi=k^2\psi. \label{difeq}\ee  With the help of  the change of
variables \be\psi(x)=e^{ikx}\phi(\xi),\qquad\xi={1\over{e^{2\alpha
x}+1}}\la{chan}\ee one can  reduce Eq. \eqref{difeq} to the Gauss
hypergeometric equation
$$\xi(\xi-1)\phi''+[(a+b+1)\xi-c]\phi'+ab\,\phi=0,$$ \be a=1-{z\over{2\alpha}},\qquad
b={z\over{2\alpha}},\qquad c=1-{{ik}\over\alpha}.\ee
Using the properties of gamma-function and hypergeometric
function of the first kind (Gauss series) $F(a,b,c;\xi)$ (see
\cite{bateman1}), one can show that for the eigenvalue $k^2$ of the Hamiltonian $h_\alpha$
there are two eigenfunctions\footnote{The asymptotics in \gl{p+'} and
\gl{p-'} are valid in the case ${\rm{Re}}\,\alpha>0$ only.},
$$\psi_+(x;k,\alpha)\!=\!{1\over\sqrt{2\pi}}\,{{\Gamma(1\!+\!{z\over{2\alpha}}
\!-\! {{ik}\over{\alpha}})
\Gamma(1\!-\!{z\over{2\alpha}}\!-\!{{ik}\over{\alpha}})}\over{\Gamma^2(1-{{ik}\over\alpha})
}}\,{{2k}\over{2k\!+\!iz}}\,e^{ikx}F\Big(1-{z\over{2\alpha}},
{z\over{2\alpha}},1-{{ik}\over\alpha};{1\over{e^{2\alpha
x}\!+\!1}}\Big)\!\equiv$$
$$\equiv{1\over\sqrt{2\pi}}\Big\{e^{ikx}
F\Big(1-{z\over{2\alpha}},{z\over{2\alpha}},1+{{ik}\over\alpha};{1\over{e^{-2\alpha
x}+1}}\Big)-$$
$$-{{\Gamma(1\!+\!{{ik}\over\alpha})\Gamma(1\!+\!{z\over{2\alpha}}
\!-\!{{ik}\over{\alpha}})
\Gamma(1\!-\!{z\over{2\alpha}}\!-\!{{ik}\over{\alpha}})}\over{\Gamma(1-{{ik}\over\alpha})}}\,
{{{{2\alpha}\over\pi}\sinh{{\pi i
z}\over{2\alpha}}}\over{2k+iz}}\,e^{-ikx}F\Big(1-{z\over{2\alpha}},{z\over{2\alpha}},
1-{{ik}\over\alpha};{1\over{e^{-2\alpha x}+1}}\Big)\Big\}=$$
\be={1\over\sqrt{2\pi}}\begin{cases}{{\Gamma(1+{z\over{2\alpha}} -
{{ik}\over{\alpha}})
\Gamma(1-{z\over{2\alpha}}-{{ik}\over{\alpha}})}\over{\Gamma^2(1-{{ik}\over\alpha})
}}\,{{2k}\over{2k+iz}}\,e^{ikx}[1+o(1)],&x\to+\infty,\\e^{ikx}[1+o(1)]
-{{\Gamma(1+{{ik}\over\alpha})\Gamma(1+{z\over{2\alpha}}
-{{ik}\over{\alpha}})
\Gamma(1-{z\over{2\alpha}}-{{ik}\over{\alpha}})}\over{\Gamma(1-{{ik}\over\alpha})}}\,
{{{{2\alpha}\over\pi}\,{\sinh}{{\pi i
z}\over{2\alpha}}}\over{2k+iz}}\,e^{-ikx}[1+o(1)],&x\to-\infty,\end{cases}\la{p+'}\ee
describing scattering for $k>0$, and
$$\psi_-(x;k,\alpha)={1\over\sqrt{2\pi}}\Big\{-{{{\rm{sinh}}\,{{\pi iz}\over{2\alpha}}}
\over{{\rm{sinh}}\,{{\pi k}\over{\alpha}}}}\,e^{ikx}
F\Big(1-{z\over{2\alpha}},{z\over{2\alpha}},
1-{{ik}\over\alpha};{1\over{e^{2\alpha x}+1}}\Big)+$$
$$+{{\Gamma^2(1-{{{ik}\over\alpha}})}\over
{\Gamma(1+{z\over{2\alpha}}-{{ik}\over\alpha})
\Gamma(1-{z\over{2\alpha}}-{{ik}\over\alpha})}}
\,\Big(1+{{iz}\over{2k}}\Big)\,e^{-ikx}F\Big(1-{z\over{2\alpha}},
{z\over{2\alpha}},1+{{ik}\over\alpha};{1\over{e^{2\alpha
x}+1}}\Big)\Big\}\equiv$$ $$\equiv
{1\over\sqrt{2\pi}}\,e^{-ikx}
F\Big(1-{z\over{2\alpha}},{z\over{2\alpha}},1-{{ik}\over\alpha};{1\over{e^{-2\alpha
x}+1}}\Big)=$$
\be={1\over\sqrt{2\pi}}\begin{cases}{{\Gamma^2(1-{{{ik}\over\alpha}})}\over
{\Gamma(1+{z\over{2\alpha}}-{{ik}\over\alpha})
\Gamma(1-{z\over{2\alpha}}-{{ik}\over\alpha})}}
\,(1+{{iz}\over{2k}}) \,e^{-ikx}[1+o(1)]-{{{\sinh}{{\pi
iz}\over{2\alpha}}}
\over{{\sinh}{{\pi k}\over{\alpha}}}}\,e^{ikx}[1\!+\!o(1)],&x\to+\infty,\\
e^{-ikx}[1+o(1)],&x\to-\infty,\end{cases}\la{p-'}\ee
describing scattering in the opposite direction for $k>0$.
These eigenfunctions are interconnected by the relation\footnote{We remind again that the standard "normalization" for scattering
is respected by $\psi_+$ where one can read off $T$ and $R$ ( transmission and reflection coefficients) but not by $\psi_-$ which in order to read off T and R would
require to be divided by $(1+iz/2k)$.}
\be\psi_-(-x;k,\alpha)\equiv{{\Gamma^2(1-{{{ik}\over\alpha}})}\over
{\Gamma(1+{z\over{2\alpha}}-{{ik}\over\alpha})
\Gamma(1-{z\over{2\alpha}}-{{ik}\over\alpha})}}
\,\Big(1+{{iz}\over{2k}}\Big)\psi_+(x;k,\alpha)\ee (cf. with
\gl{pp1}).

Taking into account properties of gamma-function and hypergeometric
function \cite{bateman1} it is not hard to check that:

\renewcommand{\labelenumi}{\rm{(\theenumi)}}
\begin{enumerate}

\item the following expressions are valid for the Wronskian of the
functions $\psi_+(x;k,\alpha)$ and $\psi_-(x;k,\alpha)$ and for
Green function:
$$W[\psi_+(x;k,\alpha),\psi_-(x;k,\alpha)]\equiv$$ \be\psi'_+(x;k,\alpha)
\psi_-(x;k,\alpha)-\psi_+(x;k,\alpha)\psi'_-(x;k,\alpha)=
{{ik}\over\pi},\ee
$$G(x,x';\lambda,\alpha)={{\pi i}\over{\sqrt\lambda}}\,\psi_+(x_>;\sqrt\lambda,\alpha)
\psi_-(x_<;\sqrt\lambda,\alpha),\qquad{\rm{Im}}\,
\sqrt\lambda\geqslant0,$$ \be(h-\lambda)G=\delta(x-x'),\qquad
x_>=\max\{x,x'\},\quad x_<=\min\{x,x'\};\la{gf43}\ee

\item in the limit for ${\rm{Re}}\,\alpha\to+\infty$ the
eigenfunctions $\psi_+(x;k,\alpha)$ and $\psi_-(x;k,\alpha)$ of the
Hamiltonian $h_\alpha$ turn into the eigenfunctions $\psi_+(x;k)$
and $\psi_-(x;k)$ respectively (see \gl{p+} and \gl{p-}) of the
Hamiltonian $h$:
\be\lim_{{\rm{Re}}\,\alpha\to+\infty}\psi_+(x;k,\alpha)=\psi_+(x;k),
\qquad\lim_{{\rm{Re}}\,\alpha\to+\infty}\psi_-(x;k,\alpha)=\psi_-(x;k);\la{lim36}\ee
hence, the Green function \gl{gf43} in the limit for
${\rm{Re}}\,\alpha\to+\infty$ becomes the Green function \gl{gf};

\item in the case ${\rm{Re}}\,\alpha\ne0$ there is spectral singularity in the
spectrum of $h_\alpha$ for the eigenvalue $E=-z^2/4$ which is
the only pole of the Green function \gl{gf43} inside of the
continuous spectrum of $h_\alpha$ and the corresponding
eigenfunction takes the form
$$\psi_0(x;\alpha)\equiv[2\,{\rm{cosh}}\,\alpha x]^{z/(2\alpha)}=\varphi^-_\alpha(x)=$$
\be-\sqrt{2\pi}\,{{\Gamma^2(1-{z\over{2\alpha}})}
\over{\Gamma(1-{z\over\alpha})}}\lim_{k\to-iz/2}[(1+{{2k}\over{iz}})\psi_+(x;k,\alpha)]=
\sqrt{2\pi}\,\psi_-(x;-iz/2,\alpha).\la{fun36}\ee

\end{enumerate}

With the help of the Green function method one can construct  the
resolution of identity from the eigenfunctions \gl{p+'} and
\gl{p-'}, \be\delta(x-x')=\int_{\cal
L}\psi_+(x;k,\alpha)\psi_-(x';k,\alpha)\,dk,\la{rez43}\ee where
$\cal L$ is the same path as in \gl{del}. Taking into account
\gl{fun36} it is not hard to rearrange \gl{rez43} to the form
\be\delta(x-x')={\rm{p.v.}}\int\limits_{-\infty}^{+\infty}\psi_+(x;k,\alpha)
\psi_-(x';k,\alpha)\,dk-{z\over4}\,{{\Gamma(1-{z\over\alpha})}
\over{\Gamma^2(1-{z\over{2\alpha}})}}\,\psi_0(x;\alpha)\psi_0(x';\alpha),\ee
which is analogous to \gl{rez26} and evidently converts into
\gl{rez26} in the limit for ${\rm{Re}}\,\alpha\to+\infty$.

Let us note at last that after the change of variable $
\tau=\tanh\alpha x$ Eq. \gl{difeq} takes the form of the Legendre
equation
\be(1-\tau^2){{d^2\psi}\over{d\tau^2}}-2\tau{{d\psi}\over{d\tau}}+\Big[\nu(\nu+1)-
{\mu^2\over{1-\tau^2}}\Big]\psi=0\ee with
\be\mu={{ik}\over\alpha},\qquad\nu=-{z\over{2\alpha}}.\ee Hence, the
eigenfunctions $\psi_+(x;k,\alpha)$ and $\psi_-(x;k,\alpha)$ can be
expressed through modified Legendre functions \cite{bateman1}. Using
\gl{p+'}, \gl{p-'} and the relation 3.4.6 from \cite{bateman1} one
can receive the following presentations for these eigenfunctions:
$$\psi_+(x;k,\alpha)={1\over\sqrt{2\pi}}\,{{\Gamma(1+{z\over{2\alpha}}
-{{ik}\over{\alpha}})
\Gamma(1-{z\over{2\alpha}}-{{ik}\over{\alpha}})}\over{\Gamma(1-{{ik}\over\alpha})
}}\,{{2k}\over{2k+iz}}\,{\rm
P}^{ik/\alpha}_{-z/(2\alpha)}(\tanh\alpha x),$$
\be\psi_-(x;k,\alpha)=
{1\over\sqrt{2\pi}}\,\Gamma\Big(1-{{ik}\over\alpha}\Big){\rm
P}^{ik/\alpha}_{-z/(2\alpha)}(-\tanh\alpha x), \label{legen}\ee where ${\rm
P}^{\mu}_{\nu}(\tau)$ is modified associated Legendre function of
the first kind.

After the change $z\to-z$ the formulae \gl{gam'} -- \gl{legen} are
valid as well for the Ha\-mil\-to\-ni\-an~$h^-_\alpha$.  These
formulae are valid as well, generally speaking, for complex $k$ and $z$ (in particular,
for real $z$) and
for  purely imaginary $\alpha$. In the latter case it is better to shift
$x\to x-x_0$, ${\rm{Im}}\,x_0\ne0$ in order to have a nonsingular
complex periodic potential. In the case
${\rm{Re}}\,z\,{\rm{Re}}\,\alpha<0$ the function \gl{fun36}  is a
wave function of the bound state of $h_\alpha$ for the energy level
$E=-z^2/4$. It follows from \gl{p+'} and \gl{p-'} that the potential
of $h_\alpha$ is reflectionless iff ${\rm{Re}}\,\alpha\ne0$,
$z=2\alpha n$, $n=0$, $\pm1$, $\pm2$,\,~\dots\,.

\section{Conclusions}

We have proven that wave functions related to spectral singularities are quite relevant to build the complete resolution of identity for non-Hermitian systems. Remarkably, depending on the class of test functions the structure of a resolution of identity  is different in the sense of distribution theory. Nevertheless, there is no class of physically motivated test functions (wave packets) for which spectral singularities are negligible.
Thus they are physical in the case when the system is characterized by a
non-Hermitian potential. Namely, they contribute to transmission and reflection coefficients dramatically enhancing their values.

We have studied in detail the delta-like and a related smooth potential. But let us notice that one can exhibit additional smooth superpotentials,
whose limiting cases coincide with $\chi(x)$ from \gl{sp22}, namely:
$$\chi_\varepsilon(x)\equiv{{zx}\over{2\sqrt{x^2+\varepsilon^2}}}\to\chi(x)\equiv
{z\over2}\,{\rm{sign}}\,x,\qquad\varepsilon\to0,$$
\be\chi_\varepsilon(x)\equiv{{z}\over{\pi}}\,{\rm{arctg}}\,{x\over\varepsilon}\to
\chi(x)\equiv{z\over2}\,{\rm{sign}}\,x,\qquad\varepsilon\downarrow0.\ee

\section*{Acknowledgments}

The work of AA and AS was supported by grant RFBR 09-01-00145-a. AA
was supported in part by CUR Generalitat de Catalunya
under project 2009SGR502 .

\section*{{APPENDIX 1.}\\ Proof of biorthogonality relations}

In order to prove \gl{dk} we shall show that
\be\lim_{A\to+\infty}\int\limits_{-\infty}^{+\infty}\Big\{\int\limits_{-A}^A
[(1+{{2k}\over{iz}})\psi_+(x;k)]\psi_-(x;k') \,dx
\Big\}\varphi(k)\,dk=(1+{{2k'}\over{iz}})\varphi(k') \la{lim1}\ee
for any test function $\varphi(k)\in C^\infty_{\Bbb R}\cap L_1(\Bbb
R)\cap\{f(k):f'(k)\in L_1(\Bbb R)\}$. It follows
from the condition $\varphi(k)\in C^\infty_{\Bbb R}\cap L_1(\Bbb
R)\cap\{f(k):f'(k)\in L_1(\Bbb R)\}$ and the relation
$$\varphi(x)=\varphi(x_0)+\int\limits_{x_0}^x\varphi'(t)\,dt$$ that
\be\lim_{x\to+\infty}\varphi(x)=\lim_{x\to-\infty}\varphi(x)=0.\la{fpm}\ee
With the help of  a straightforward calculation one can transform the
left-hand part of \gl{lim1} to
\be{1\over\pi}\lim_{A\to+\infty}\int\limits_{-\infty}^{+\infty}\Big\{
{{\sin A(k-k')}\over{k-k'}}\,\psi(k)-{{\sin k'A}\over
k'}\,e^{ikA}\varphi(k)
\Big\}\,dk,\qquad\psi(k)=(1+{{2k}\over{iz}})\varphi(k), \label{int1}\ee where in the case $k'=0$ the
ratio $\sin k'A/k'$  is equal to $A$. It follows
from the conditions \gl{fpm}, the inclusion $\varphi(x)\in L_1(\Bbb
R)\cap\{f(k):f'(k)\in L_1(\Bbb R)\}$ and the Riemann theorem that
\be\lim_{A\to+\infty}\Big\{{{\sin k'A}\over
k'}\int\limits_{-\infty}^{+\infty} \,e^{ikA}\varphi(k)\,dk
\Big\}=\lim_{A\to+\infty}\Big\{i{{\sin k'A}\over
k'A}\int\limits_{-\infty}^{+\infty} \,e^{ikA}\varphi'(k)\,dk
\Big\}=0.\ee Thus in the limit $A\to+\infty$ the expression \eqref{int1} is reduced  to the first term. By virtue of the Riemann
theorem and due to the evident inclusions
$${{\psi(k)}\over{k-k'}}\in L_1(\Bbb R\backslash]k'-\delta,k'+\delta[),\qquad
{{\psi(k)-\psi(k')}\over{k-k'}}\in L_1([k'-\delta,k'+\delta])$$ for
any $\delta>0$, the following relations are valid:
$$\lim_{A\to+\infty}\Big(\int\limits_{-\infty}^{k'-\delta}+
\int\limits_{k'+\delta}^{+\infty}\Big)\sin
A(k-k'){{\psi(x)}\over{k-k'}}\,dk=0,$$ \be\lim_{A\to+\infty}
\int\limits_{k'-\delta}^{k'+\delta}\sin
A(k-k'){{\psi(k)-\psi(k')}\over{k-k'}}\,dk=0.\ee Hence,
\be{1\over\pi}\lim_{A\to+\infty}\int\limits_{-\infty}^{+\infty}
{{\sin A(k-k')}\over{k-k'}}\,\psi(k)\,dk={{\psi(k')}\over{\pi}}
\lim_{A\to+\infty}\int\limits_{k'-\delta}^{k'+\delta} {{\sin
A(k-k')}\over{k-k'}}\,dk=\psi(k').\ee Thus, \gl{lim1} and,
consequently, \gl{dk} are valid.

Let us notice that one can prove \gl{dk} also for test functions from a wider class with the help of the technique of Theorem 2 and
Remark 1 of the next Appendix 2.

\section*{APPENDIX 2.\\ Proofs of resolutions of identity}

Let $CL_\gamma=C_{\Bbb R}\cap\{f(x):f(x)\big|_{(-\infty,0]}\in
C^\infty_{(-\infty,0]}, f(x)\big|_{[0,+\infty)}\in
C^\infty_{[0,+\infty)}\}\cap L_2(\Bbb R;(1+|x|)^\gamma)$,
$\gamma\in\Bbb R$, be the space of test functions.\footnote{Such a
choice of the test functions space is motivated by properties of $h$
eigenfunctions (see \gl{p+} and \gl{p-}).} The sequence
$\varphi_n(x)\in CL_\gamma$ is called convergent in $CL_\gamma$ to
$\varphi(x)\in CL_\gamma$,
\be\mathop{\lim\nolimits_\gamma}_{n\to+\infty}\varphi_n(x)=\varphi(x)\ee
if
\be\lim_{n\to+\infty}\int\limits_{-\infty}^{+\infty}|\varphi_n(x)-\varphi(x)|^2
(1+|x|)^\gamma dx=0,\ee and for any $x_1, x_2\in \Bbb R$, $x_1<x_2$,
\be\lim_{n\to+\infty}\max_{[x_1,x_2]}|\varphi_n(x)-\varphi(x)|=0.\ee

We shall denote the value of a functional $f$ on $\varphi\in
CL_\gamma$ conventionally as $(f,\varphi)$. A functional $f$ is
called continuous if for any sequence $\varphi_n\in CL_\gamma$
convergent in $CL_\gamma$ to zero the equality
\be\lim_{n\to+\infty}(f,\varphi_n)=0\ee is valid. The space of
distributions over $CL_\gamma$, {\it i.e.} of linear continuous
functionals over $CL_\gamma$ is denoted $CL'_\gamma$. The sequence
$f_n\in CL'_\gamma$ is called convergent in $CL'_\gamma$ to $f\in
CL'_\gamma$,
\be\mathop{\lim\nolimits'_\gamma}_{n\to+\infty}f_n=f,\ee if for any
$\varphi \in CL_\gamma$ the relation takes place,
\be\lim_{n\to+\infty}(f_n,\varphi)=(f,\varphi).\ee

A functional $f\in CL'_\gamma$ is called regular if there is
$f(x)\in L_2(\Bbb R;(1+|x|)^{-\gamma})$ such that for any
$\varphi\in CL_\gamma$ the equality
\be(f,\varphi)=\int\limits_{-\infty}^{+\infty}f(x)\varphi(x)\,dx\ee
holds. In this case we shall identify the distribution $f\in
CL'_\gamma$ with the function $f(x)\in L_2(\Bbb
R;(1+|x|)^{-\gamma})$. In virtue of Bunyakovskii inequality,
\be\Big|\int\limits_{-\infty}^{+\infty}f(x)\varphi(x)\,dx\Big|^2\leqslant
\int\limits_{-\infty}^{+\infty}{{|f^2(x)|\,dx}\over{(1+|x|)^\gamma}}
\int\limits_{-\infty}^{+\infty}|\varphi^2(x)|(1+|x|)^\gamma\,dx,\ee
it is evident that $L_2(\Bbb R;(1+|x|)^{-\gamma})\subset CL'_\gamma$
and this inclusion is continuous.

For any $\gamma_1<\gamma_2$ there is obviously continuous inclusion
$CL_{\gamma_2}\subset CL_{\gamma_1}$.
Let us also note that the Dirac delta function $\delta(x-x')$
belongs to $CL'_\gamma$ for any $\gamma\in\Bbb R$.

Proof of the resolution of identity \gl{del} is based on the following
lemma.\\

\noindent {\bf Lemma 1.} {\it Suppose that ${\cal L}(A)$ is a path
in complex $k$ plane , made of the segment $[-A,A]$ by its
deformation near the point $k=k_0$, $k_0\in(-A,A)\subset\Bbb R$
upwards and the direction of ${\cal L}(A)$ is specified from
$-A$ to $A$. Then for any $r>0$, $k_0\in\Bbb R$ and $A>|k_0|$ the
inequalities hold \be\Big|\int_{{\cal
L}(A)}{{e^{ikr}}\over{k-k_0}}\,dk\,\Big|\leqslant{{AD}\over{(1+r(A-|k_0|))
(A-|k_0|)}}\la{in34}\ee and \be\Big|\int_{{\cal
L}(A)}{{e^{ikr}}\over{k-k_0}}\,dk-{1\over{ir}}\Big({e^{iAr}\over{A-k_0}}+
{e^{-iAr}\over{A+k_0}}\Big)\,\Big|\leqslant{4\over{(A-|k_0|)^2r^2}},\la{ner40}\ee
where $D>0$ is a constant independent of $r$, $k_0$ and
$A$.}\\

\noindent{\bf Proof.} With the help of Jordan lemma one can easily
check that
$$\int_{{\cal
L}(A)}{{e^{ikr}}\over{k-k_0}}\,dk=-\Big(\int\limits_{-\infty}^{-A}+
\int\limits_A^{+\infty}\Big){{e^{ikr}}\over{k-k_0}}\,dk=
\int\limits_{A+k_0}^{+\infty}e^{ik_0r-i\tau
r}\,{{d\tau}\over{\tau}}-\int\limits_{A-k_0}^{+\infty}e^{ik_0r+i\tau
r}\,{{d\tau}\over{\tau}}=$$ \be
e^{ik_0r}\Big(\int\limits_{A+k_0}^{A-k_0}{{\cos\tau
r}\over{\tau}}\,d\tau-i\int\limits_{A+k_0}^{+\infty}{{\sin\tau
r}\over{\tau}}\,d\tau-i\int\limits_{A-k_0}^{+\infty}{{\sin\tau
r}\over{\tau}}\,d\tau\Big).\la{int35}\ee For the first integral in
the right-hand side of \gl{int35} the following estimation is valid:
$$\Big|\int\limits_{A+k_0}^{A-k_0}{{\cos\tau
r}\over{\tau}}\,d\tau\Big|=\Big|{{\sin
r(A-k_0)}\over{r(A-k_0)}}-{{\sin
r(A+k_0)}\over{r(A+k_0)}}+\int\limits_{A+k_0}^{A-k_0}{{\sin\tau
r}\over{r\tau^2}}\,d\tau\Big|\leqslant$$
$${{C/2}\over{1+r(A-k_0)}}+{{C/2}\over{1+r(A+k_0)}}+\int\limits_{A-|k_0|}^{A+|k_0|}
{{C/2}\over{\tau(1+r\tau)}}\,d\tau\leqslant{C\over{1+r(A-|k_0|)}}+$$
$${C\over2}\ln{{1+1/[r(A-|k_0|)]}\over{1+1/[r(A+|k_0|)]}}\leqslant
{C\over{1+r(A-|k_0|)}}+{C\over2}\Big\{{{1+1/[r(A-|k_0|)]}\over{1+1/[r(A+|k_0|)]}}-1\Big\}=$$
\be{C\over{1+r(A-|k_0|)}}+{{C|k_0|}\over{(1+r(A+|k_0|))(A-|k_0|)}}\leqslant
{{AC}\over{(1+r(A-|k_0|))(A-|k_0|)}},\ee where
$C=2\sup_{\xi>0}|(1+1/\xi)\sin\xi|$ and the obvious inequality
$\ln\xi\leqslant\xi-1$, $\xi\geqslant0$ is used. For the second and
third integrals of the right-hand side of \gl{int35} we have:
$$\Big|\int\limits_{A\pm k_0}^{+\infty}{{\sin r\tau}\over{\tau}}\,d\tau\Big|=
{2\over r}\Big|\int\limits_{A\pm k_0}^{+\infty}{{d\sin^2
(r\tau/2)}\over{\tau}}\Big|={2\over r}\Big|-{{\sin^2[r(A\pm
k_0)/2]}\over{A\pm k_0}}+\int\limits_{A\pm k_0}^{+\infty}{{\sin^2
(r\tau/2)}\over{\tau^2}}\,d\tau\Big|\leqslant$$ $${{(C/2)^2r(A\pm
k_0)/2}\over{[1+r(A\pm k_0)/2]^2}}+\int\limits_{A\pm
k_0}^{+\infty}{{(C/2)^2r/2}\over{(1+r\tau/2)^2}}\,d\tau=
{{(C/2)^2r(A\pm k_0)/2}\over{[1+r(A\pm
k_0)/2]^2}}+{{(C/2)^2}\over{1+r(A\pm k_0)/2}}\leqslant$$
\be{{C^2/2}\over{1+r(A-|k_0|)/2}}\leqslant{{C^2}\over{1+r(A-|k_0|)}}\leqslant
{{AC^2}\over{(1+r(A-|k_0|))(A-|k_0|)}}.\la{ner37}\ee The inequality
\gl{in34} follows from \gl{int35} -- \gl{ner37} with $D=C+2C^2$.

The inequality \gl{ner40} is valid in view of the following chain of relations
derived with the help of \gl{int35} and integration by parts:
$$\Big|\int_{{\cal
L}(A)}{{e^{ikr}}\over{k-k_0}}\,dk-{1\over{ir}}\Big({e^{iAr}\over{A-k_0}}+
{e^{-iAr}\over{A+k_0}}\Big)\,\Big|=$$ $${1\over
r}\Big|\Big(\int\limits_{-\infty}^{-A}+
\int\limits_A^{+\infty}\Big){{e^{ikr}}\over{(k-k_0)^2}}\,dk\Big|=
{1\over
r^2}\Big|{e^{-iAr}\over{(A+k_0)^2}}-{e^{iAr}\over{(A-k_0)^2}}+
2\Big(\int\limits_{-\infty}^{-A}+
\int\limits_A^{+\infty}\Big){{e^{ikr}}\over{(k-k_0)^3}}\,dk\Big|\leqslant$$
\be{2\over{(A-|k_0|)^2r^2}}+{4\over
r^2}\int\limits_A^{+\infty}{{dk}\over{(k-|k_0|)^3}}={4\over{(A-|k_0|)^2r^2}}.\ee
Lemma 1 is proved.\\

Validity of the resolution of identity \gl{del} in $CL'_\gamma$ for any
$\gamma>-1$ is a corollary of the following theorem.\\

\noindent{\bf Theorem 1.} {\it Suppose that

\renewcommand{\labelenumi}{\rm{(\theenumi)}}
\begin{enumerate}

\item the functions $\psi_+(x;k)$ and $\psi_-(x;k)$ are defined by
the formulae \gl{p+} and \gl{p-} respectively for any $x\in\Bbb R$,
$k\in\Bbb C$ and fixed purely imaginary $z\ne0$;

\item ${\cal L}(A)$ is a path in complex $k$ plane , made of the
segment $[-A,A]$ by its deformation near the point $k=-iz/2$ upwards
and the direction of ${\cal L}(A)$ is specified from $-A$
to $A$.

\end{enumerate}

\noindent Then for any $\gamma>-1$ and $x'\in\Bbb R$ the following
relation holds: \be\mathop{{\lim}'_\gamma}_{A\to+\infty}\int_{{\cal
L}(A)}\psi_+(x;k)\psi_-(x';k)\,dk=\delta(x-x').\la{del38}\ee}\\

\noindent {\bf Proof.} One can reduce the product
$\psi_+(x;k)\psi_-(x';k)$  to the form
\be\psi_+(x;k)\psi_-(x';k)={1\over{2\pi}}\Big\{e^{ik(x-x')}+{{2iz}\over
k}\,\theta(-x)\theta(x')\sin kx\,\sin kx'
-{{iz}\over{2k+iz}}\,e^{ik(|x|+|x'|)}\Big\},\la{pred39}\ee where
notation \gl{not4} is used. Hence, \be\int_{{\cal
L}(A)}\psi_+(x;k)\psi_-(x';k)\,dk= {1\over\pi}{{\sin
A(x-x')}\over{x-x'}}-{{iz}\over{4\pi}}\int_{{\cal
L}(A)}{{e^{ik(|x|+|x'|)}}\over{k+iz/2}}\,dk.\la{int40}\ee  In view of Lemma 1 the integral
in the left-hand part of \gl{int40} belongs to
$L_2(\Bbb R;(1+|x|)^{-\gamma})$ and therefore to $CL'_\gamma$. Thus,
to prove \gl{del38} it is sufficient to establish that for any
$\varphi(x)\in CL_\gamma$ the equality takes place,
\be\lim_{A\to+\infty}{1\over\pi}\int\limits_{-\infty}^{+\infty}
\Big[{{\sin A(x-x')}\over{x-x'}}-{{iz}\over{4}}\int_{{\cal
L}(A)}{{e^{ik(|x|+|x'|)}}\over{k+iz/2}}\,dk\Big]
\varphi(x)\,dx=\varphi(x'). \label{int2}\ee By virtue of Bunyakovskii inequality,
Lemma 1 and  inequality
\be(1+r)^\beta\leqslant1+r^\beta,\qquad
r\geqslant0,\quad0\leqslant\beta\leqslant1\la{ner42}\ee we have:
$$\Big|\int\limits_{-\infty}^{+\infty} \Big[\int_{{\cal
L}(A)}{{e^{ik(|x|+|x'|)}}\over{k+iz/2}}\,dk\Big]
\varphi(x)\,dx\Big|^2\leqslant$$
$$\int\limits_{-\infty}^{+\infty}\Big|\int_{{\cal
L}(A)}{{e^{ik(|x|+|x'|)}}\over{k+iz/2}}\,dk\Big|^2{{dx}\over{(1+|x|)^\gamma}}
\int\limits_{-\infty}^{+\infty}|\varphi^2(x)|(1+|x|)^\gamma\,dx\leqslant$$
$$ {{A^2D^2}\over{(A-|z|/2)^2}}\int\limits_{-\infty}^{+ \infty}
{{1+\theta(-\gamma)|x|^{-\gamma}}\over{[1+(|x|+|x'|)(A-|z|/2)]^2}}\,dx
\int\limits_{-\infty}^{+\infty}|\varphi^2(x)|(1+|x|)^\gamma\,dx\leqslant$$
$${{2A^2D^2}\over{(A-|z|/2)^2}}\int\limits_0^{+ \infty}
{{1+\theta(-\gamma)x^{-\gamma}} \over{[1+x(A-|z|/2)]^2}}\,dx
\int\limits_{-\infty}^{+\infty}|\varphi^2(x)|(1+|x|)^\gamma\,dx=$$
\be{{2A^2D^2}\over{(A\!-\!|z|/2)^3}}\!\int\limits_0^{+ \infty}
{{1\!+\!\theta(-\gamma)(A\!-\!|z|/2)^\gamma\xi^{-\gamma}}
\over{(1+\xi)^2}}\,d\xi
\!\int\limits_{-\infty}^{+\infty}\!|\varphi^2(x)|(1+|x|)^\gamma\,dx\to0,\qquad
A\to+\infty,\ee wherefrom it follows that the second term in the left-hand part of \eqref{int2} vanishes. Now the proof that the left-hand side of \gl{int2} is equal to
$\varphi(x')$ is analogous to the proof in Appendix 1. Thus, Theorem 1
is proved.\\

The applicability of the resolution of identity \gl{del} for some bounded and
slowly increasing test functions is based on the next
theorem.\\

\noindent{\bf Theorem 2.} {\it Suppose that

\renewcommand{\labelenumi}{\rm{(\theenumi)}}
\begin{enumerate}

\item the functions $\psi_+(x;k)$ and $\psi_-(x;k)$ are defined by
the formulae \gl{p+} and \gl{p-} respectively for any $x\in\Bbb R$,
$k\in\Bbb C$ and fixed purely imaginary $z\ne0$;

\item ${\cal L}(A)$ is a path in complex $k$ plane, made of the
segment $[-A,A]$ by its deformation near the point $k=-iz/2$ upwards
and the direction of ${\cal L}(A)$ is specified from $-A$
to $A$;

\item the function $\eta(x)\in C^\infty_{\Bbb R}$,
$\eta(x)\equiv0$ for any $x\leqslant1$, $\eta(x)\in[0,1]$ for any
$x\in[1,2]$ and $\eta(x)\equiv1$ for any $x\geqslant2$.

\end{enumerate}

\noindent Then for any $\varkappa\in[0,1)$, $k_0\in\Bbb R$ and
$x'\in\Bbb R$ the following relation holds:
\be\lim_{A\to+\infty}\int\limits_{-\infty}^{+\infty}\Big[\int_{{\cal
L}(A)}\psi_+(x;k)\psi_-(x';k)\,dk\Big]\Big[\eta(\pm
x)e^{ik_0x}|x|^\varkappa\Big]\,dx=
\eta(\pm x')e^{ik_0x'}|x'|^\varkappa.\la{int52}\ee}\\

\noindent {\bf Proof.} We present the proof for the case with upper
signs in \gl{int52} only because the proof for the case with lower
signs is quite similar. By virtue of \gl{int40} and Lemma 1 the
following asymptotics takes place for the integral over $k$ in
\gl{int52}:
$$\int_{{\cal L}(A)}\psi_+(x;k)\psi_-(x';k)\,dk= {1\over\pi}{{\sin
A(x-x')}\over{x-x'}}-$$ \be{{z}\over{4\pi}}
\Big({e^{iA(|x|+|x'|)}\over{A+iz/2}}+{e^{-iA(|x|+|x'|)}\over{A-iz/2}}
\Big){1\over{|x|+|x'|}}+O\Big({1\over x^2}\Big),\qquad
x\to\pm\infty.\la{as53}\ee Hence, the integral over $x$ in
\gl{int52} converges for any $A>|k_0|$. It follows from Theorem~1
that
$$\lim_{A\to+\infty}\int\limits_{-\infty}^{+\infty}\Big[\int_{{\cal
L}(A)}\psi_+(x;k)\psi_-(x';k)\,dk\Big]\Big[\eta(
x)\eta(3+x'-x)e^{ik_0x}|x|^\varkappa\Big]\,dx=$$ \be=\eta(
x')\eta(3)e^{ik_0x'}|x'|^\varkappa=\eta(
x')e^{ik_0x'}|x'|^\varkappa.\la{int54}\ee Hence, to prove \gl{int52}
it is sufficient to prove that
\be\lim_{A\to+\infty}\int\limits_{-\infty}^{+\infty}\Big[\int_{{\cal
L}(A)}\psi_+(x;k)\psi_-(x';k)\,dk\Big]\Big\{\eta(
x)[1-\eta(3+x'-x)]e^{ik_0x}|x|^\varkappa\Big\}\,dx=0.\la{lim55}\ee
The fact that contributions of the first and second terms of the
right-hand side of \gl{as53} vanish in the limit \gl{lim55} can be checked with the help of Riemann theorem and
integration by parts.  In view of
Lemma 1 and \gl{int40} the contribution of the third term of the
right-hand side of \gl{as53} in the integral \gl{lim55} does not exceed the following expression:
$${{|z|}\over{\pi(A-|z|/2)^2}}\int\limits_{-\infty}^{+\infty}
{{\eta( x)[1-\eta(3+x'-x)]|x|^\varkappa}\over{(|x|+|x'|)^2}}\,dx=$$
\be={{|z|}\over {\pi(A-|z|/2)^2}}\int\limits_1^{+\infty}{{\eta(
x)[1-\eta(3+x'-x)]|x|^\varkappa}\over{(|x|+|x'|)^2}}\,dx\to0,\qquad
A\to+\infty.\ee
Theorem 2 is proved.\\

\noindent{\bf Remark 1.} Theorems 1 and 2 provide the validity of
resolution of identity \gl{del} for test functions which are linear
combinations of functions $\eta(\pm x)e^{ik_0x}|x|^\varkappa$, in general, with
different  $\varkappa\in[0,1)$ and $k_0\in\Bbb R$ and
functions from $CL_\gamma$, in general, with different $\gamma>-1$. In
particular, these theorems guarantee applicability of \gl{del} for
eigenfunctions \gl{p+} and \gl{p-} of the Hamiltonian $h$.\\

\noindent{\bf Remark 2.} One can rearrange the
resolution of identity \gl{del} to the form \gl{ree15} with the help of
\gl{pred39} and the chain of relations,
$$-{{iz}\over{2\pi}}\int_{\ell(\varepsilon)}{{e^{ik(|x|+|x'|)}}
\over{2k+iz}}\,dk=$$
$$-{{iz}\over{2\pi}}\Big[e^{z(|x|+|x'|)/2}\int_{\ell
(\varepsilon)}{{dk}\over{2k+iz}}+\int_{\ell
(\varepsilon)}{{e^{ik(|x|+|x'|)}-e^{z(|x|+|x'|)/2}}
\over{2k+iz}}\,dk\Big]=$$
$$=-{{iz}\over{2\pi}}\,e^{z(|x|+|x'|)/2}\Big[-{{\pi
i}\over2}+\int\limits_{-iz/2-\varepsilon}^{-iz/2+\varepsilon}
{{e^{i(k+iz/2)(|x|+|x'|)}-1}\over{2k+iz}}\,dk\Big]=$$
$$=-{{iz}\over{2\pi}}\,\psi_0(x)\psi_0(x')\Big[-{{\pi
i}\over2}+{i\over2}\int\limits_{-iz/2-\varepsilon}^{-iz/2+\varepsilon}
{{\sin[(k+iz/2)(|x|+|x'|)]}\over{k+iz/2}}\,dk\Big]=$$
\be=-{{z}\over{4}}\,\psi_0(x)\psi_0(x')\Big[1-{2\over\pi}
\int\limits_0^{\varepsilon(|x|+|x'|)} {{\sin t}\over t}\,dt\Big]
\equiv-{{z}\over{2\pi}}\,\psi_0(x)\psi_0(x')
\int\limits_{\varepsilon(|x|+|x'|)}^{+\infty} {{\sin t}\over
t}\,dt,\ee where the relation 2.5.3.12 from \cite{prud} is taken
into account and $\ell(\varepsilon)$ is the integration path in
complex plane of $k$ defined by the equation $k=-iz/2+\varepsilon
e^{i(\pi-\vartheta)}$, $0\leqslant\vartheta\leqslant\pi$ with the
direction corresponding to
increasing $\vartheta$.\\

The resolutions of identity \gl{re16} and \gl{re15} are based on the
following lemmas.\\

\noindent{\bf Lemma 2.} {\it For any $\gamma>-1$ and $x'\in\Bbb R$
the relation holds
\be\mathop{{\lim}'_\gamma}\limits_{\varepsilon\downarrow0}
{{\sin\varepsilon(x-x')}\over{x-x'}}=0 .\ee}\\

\noindent{\bf Proof.} It is true that
\be{{\sin\varepsilon(x-x')}\over{x-x'}}\in L_2(\Bbb
R;(1+|x|)^{-\gamma})\subset CL'_\gamma,\qquad\gamma>-1.\ee Thus, to
prove the lemma it is sufficient to establish that for any
$\varphi(x)\in CL_\gamma$, $\gamma>-1$, the relation
\be\lim_{\varepsilon\downarrow0}
\int\limits_{-\infty}^{+\infty}{{\sin\varepsilon(x-x')}\over{x-x'}}
\,\varphi(x)\,dx=0\ee is valid. But its validity follows from
Bunyakovskii inequality and \gl{ner42}:
$$\Big|\int\limits_{-\infty}^{+\infty}{{\sin\varepsilon(x-x')}\over{x-x'}}
\,\varphi(x)\,dx\Big|^2\leqslant\int\limits_{-\infty}^{+\infty}
{{\sin^2\varepsilon(x-x')\,dx}\over{(x-x')^2(1+|x|)^\gamma}}
\int\limits_{-\infty}^{+\infty}|\varphi^2(x)|(1+|x|)^\gamma\,dx\leqslant$$
$$\int\limits_{-\infty}^{+\infty}
{{\sin^2\varepsilon(x-x')}\over{(x-x')^2}}\,[1+\theta(-\gamma)(|x'|
+|x-x'|)^{-\gamma}] \,dx
\int\limits_{-\infty}^{+\infty}|\varphi^2(x)|(1+|x|)^\gamma\,dx\leqslant$$
$$\int\limits_{-\infty}^{+\infty}
{{\sin^2\varepsilon(x-x')}\over{(x-x')^2}}\,[1+\theta(-\gamma)(|x'|^{-\gamma}
+|x-x'|^{-\gamma})] \,dx
\int\limits_{-\infty}^{+\infty}|\varphi^2(x)|(1+|x|)^\gamma\,dx=$$
\be\varepsilon\int\limits_{-\infty}^{+\infty}
{{\sin^2\xi}\over{\xi^2}}\,[1+\theta(-\gamma)(|x'|^{-\gamma}
+\varepsilon^\gamma|\xi|^{-\gamma})] \,d\xi
\int\limits_{-\infty}^{+\infty}|\varphi^2(x)|(1+|x|)^\gamma\,dx\to0,
\qquad\varepsilon\downarrow0,\la{cep48}\ee where we define that the
value $|x'|^{-\gamma}$ for $x'=0$ and $\gamma=0$ is equal to zero.
Lemma 2 is proved.\\

\noindent{\bf Lemma 3.} {\it For any $y\in\Bbb R$, $\varepsilon>0$
and $k_0\in \Bbb R$, $|k_0|>\varepsilon$ the inequality
\be\Big|\int\limits_{k_0-\varepsilon}^{k_0+\varepsilon}{e^{iky}\over
k}\,dk\Big|\leqslant{{\varepsilon
D}\over{(|k_0|-\varepsilon)(2+\varepsilon|y|)}}\ee
takes place, where $D$ is a constant independent of $y$, $\varepsilon$ and $k_0$.}\\

\noindent{\bf Proof.} Lemma 3 is valid by virtue of the following
chain of relations:
$$\Big|\int\limits_{k_0-\varepsilon}^{k_0+\varepsilon}{e^{iky}\over
k}\,dk\Big|=\Big|e^{ik_0y}\int\limits_{-\varepsilon}^{\varepsilon}{e^{i
y\tau}\over{k_0+\tau}}\,d\tau\Big|=
\Big|{1\over{iy}}\int\limits_{-\varepsilon}^{\varepsilon}{{d(e^{i
y\tau}-1)}\over{k_0+\tau}}\Big|=$$
$$\Big|{{e^{i\varepsilon y}-1}\over{iy(k_0+\varepsilon)}}-{{e^{-i\varepsilon
y}-1}\over{iy(k_0-\varepsilon)}}+{1\over{iy}}\int\limits_{-\varepsilon}^{\varepsilon}
{{e^{iy\tau}-1}\over {(k_0+\tau)^2}}\,d\tau\Big|\leqslant$$
$$2{{|\sin(\varepsilon y/2)|}\over{|y(k_0+\varepsilon)|}}+2{{|\sin(\varepsilon
y/2)|}\over{|y(k_0-\varepsilon)|}}+{2\over{|y|}}\int\limits_{-\varepsilon}^{\varepsilon}
{{|\sin(y\tau/2)|}\over{(k_0+\tau)^2}}\,d\tau\leqslant$$
$${{2\varepsilon C}\over{(|k_0|-\varepsilon)(2+\varepsilon|y|)}}+
2C\int\limits_0^{\varepsilon} {{\tau\,d\tau}\over
{(|k_0|-\tau)^2(2+|y|\tau)}}\leqslant$$ $${{2\varepsilon
C}\over{(|k_0|-\varepsilon)(2+\varepsilon|y|)}}+ {{2\varepsilon
C}\over{2+\varepsilon|y|}}\int\limits_0^{\varepsilon} {{d\tau}\over
{(|k_0|-\tau)^2}}={{2\varepsilon
C}\over{2+\varepsilon|y|}}\Big[{1\over{|k_0|-\varepsilon}}+
{1\over{|k_0|-\varepsilon}}-{1\over{|k_0|}}\Big]=$$
\be{{2\varepsilon
C(|k_0|+\varepsilon)}\over{|k_0|(|k_0|-\varepsilon)(2+\varepsilon|y|)}}
\leqslant{{\varepsilon
D}\over{(|k_0|-\varepsilon)(2+\varepsilon|y|)}},\ee where $D=4C$,
$C=2\sup_{\xi\geqslant0}|(1+1/\xi)\sin\xi|$ .
Lemma 3 is proved.\\

\noindent{\bf Corollary 1.} For any $x\in\Bbb R$, $x'\in\Bbb R$ and
purely imaginary $z\ne0$ the inequality
\be\Big|\int\limits_{-iz/2-\varepsilon}^{-iz/2+\varepsilon}{1\over
k}\,\sin kx\,\sin kx'\,dk\Big|\leqslant{{\varepsilon
D}\over{(|z|/2-\varepsilon)(2+\varepsilon||x|-|x'||)}}\ee is valid.\\

\noindent{\bf Corollary 2.} In conditions of Lemma 3 the inequality
\be\Big|\int\limits_{k_0-\varepsilon}^{k_0+\varepsilon}{e^{iky}\over
{k+k_0}}\,dk\Big|=\Big|e^{-ik_0y}
\int\limits_{2k_0-\varepsilon}^{2k_0+\varepsilon}{e^{iy\tau}\over
\tau}\,d\tau\Big|=\Big|
\int\limits_{2k_0-\varepsilon}^{2k_0+\varepsilon}{e^{iy\tau}\over
\tau}\,d\tau\Big|\leqslant{{\varepsilon
D}\over{(2|k_0|-\varepsilon)(2+\varepsilon|y|)}}\ee holds as well.\\

\noindent{\bf Lemma 4.} {\it For any $\gamma>-1$, $x'\in\Bbb R$ and
purely imaginary $z\ne0$ the relation \be\mathop{{\lim}'_\gamma}
\limits_{\varepsilon\downarrow0}\Big[\theta(-x)\theta(x')
\int\limits_{-iz/2-\varepsilon}^{-iz/2+\varepsilon}{1\over k}\,\sin
kx\,\sin
kx'\,dk\Big]=0\ee takes place.}\\

\noindent{\bf Proof.} In view of Corollary 1 \be\theta(-x)\theta(x')
\int\limits_{-iz/2-\varepsilon}^{-iz/2+\varepsilon}{1\over k}\,\sin
kx\,\sin kx'\,dk\in L_2(\Bbb R;(1+|x|)^{-\gamma})\subset
CL'_\gamma,\qquad\gamma>-1.\ee Thus, to prove the lemma it is
sufficient to establish that for any $\varphi(x)\in CL_\gamma$,
$\gamma>-1$, the relation \be\lim_{\varepsilon\downarrow0}
\int\limits_{-\infty}^{+\infty}\Big[\theta(-x)\theta(x')
\int\limits_{-iz/2-\varepsilon}^{-iz/2+\varepsilon}{1\over k}\,\sin
kx\,\sin kx'\,dk\Big]\,\varphi(x)\,dx=0\ee is valid. But its
validity follows from Bunyakovskii inequality, Corollary 1 and
\gl{ner42}:
$$\Big|\int\limits_{-\infty}^{+\infty}\Big[\theta(-x)\theta(x')
\int\limits_{-iz/2-\varepsilon}^{-iz/2+\varepsilon}{1\over k}\,\sin
kx\,\sin kx'\,dk\Big]\,\varphi(x)\,dx\Big|^2\leqslant$$
$$\theta(x')\int\limits_{-\infty}^0\Big|
\int\limits_{-iz/2-\varepsilon}^{-iz/2+\varepsilon}{1\over k}\,\sin
kx\,\sin kx'\,dk\Big|^2 {{dx}\over{(1+|x|)^\gamma}}
\int\limits_{-\infty}^0|\varphi^2(x)|(1+|x|)^\gamma\,dx\leqslant$$
$${{\theta(x')\varepsilon^2D^2}\over{(|z|/2-\varepsilon)^2}}\int\limits_{-\infty}^0
{{dx}\over{(2+\varepsilon||x|-|x'||)^2(1+|x|)^\gamma}}
\int\limits_{-\infty}^0|\varphi^2(x)|(1+|x|)^\gamma\,dx=$$
$${{\theta(x')\varepsilon^2D^2}\over{(|z|/2-\varepsilon)^2}}\int\limits_0^{+\infty}
{{dx}\over{(2+\varepsilon|x-|x'||)^2(1+|x|)^\gamma}}
\int\limits_{-\infty}^0|\varphi^2(x)|(1+|x|)^\gamma\,dx\leqslant$$
$${{\theta(x')\varepsilon^2D^2}\over{(|z|/2-\varepsilon)^2}}\int\limits_{-\infty}^{+\infty}
{{d\tau}\over{(2+\varepsilon|\tau|)^2(1+|\tau+|x'||)^\gamma}}
\int\limits_{-\infty}^0|\varphi^2(x)|(1+|x|)^\gamma\,dx\leqslant$$
$${{\theta(x')\varepsilon^2D^2}\over{(|z|/2-\varepsilon)^2}}\int\limits_{-\infty}^{+\infty}
{{1+\theta(-\gamma)|\tau+|x'||^{-\gamma}}\over{(2+\varepsilon|\tau|)^2}}\,d\tau
\int\limits_{-\infty}^0|\varphi^2(x)|(1+|x|)^\gamma\,dx\leqslant$$
\be{{\theta(x')\varepsilon
D^2}\over{(|z|/2-\varepsilon)^2}}\int\limits_{-\infty}^{+\infty}
{{1+\theta(-\gamma)(|x'|^{-\gamma}+\varepsilon^\gamma|\xi|^{-\gamma})}
\over{(2+|\xi|)^2}}\,d\xi
\int\limits_{-\infty}^0|\varphi^2(x)|(1+|x|)^\gamma\,dx\to0,
\qquad\varepsilon\downarrow0,\la{cep48}\ee where we define that the
value $|x'|^{-\gamma}$ for $x'=0$ and $\gamma=0$ is equal to zero.
Lemma 4 is proved.\\

{\bf Remark 3.} Let us consider the functionals
\be\mathop{{\lim}''_\gamma}_{\varepsilon\downarrow0}
{{\sin\varepsilon(x-x')}\over{x-x'}},\qquad\mathop{{\lim}''_\gamma}_{
\varepsilon\downarrow0}\Big[\theta(-x)\theta(x')
\int\limits_{-iz/2-\varepsilon}^{-iz/2+\varepsilon}{1\over k}\,\sin
kx\,\sin kx'\,dk\Big],\la{fun96}\ee which are defined by the
expressions
$$\lim_{\varepsilon\downarrow0}\int\limits_{-\infty}^{+\infty}
{{\sin\varepsilon(x-x')}\over{x-x'}}\,\varphi(x)\,dx,$$
\be\lim_{\varepsilon\downarrow0}\int\limits_{-\infty}^{+\infty}\Big[\theta(-x)\theta(x')
\int\limits_{-iz/2-\varepsilon}^{-iz/2+\varepsilon}{1\over k}\,\sin
kx\,\sin kx'\,dk\Big]\varphi(x)\,dx,\la{lim97}\ee for all test
functions $\varphi(x)\in CL_\gamma$, $\gamma\in\Bbb R$, for which
corresponding to \gl{fun96} limits from \gl{lim97} exist. It follows
from Lemmas 2 and 4 that these functionals are trivial (equal to
zero) for any $\gamma>-1$, but at the same time in view of
\gl{cep20} and \gl{cep21} these functionals are nontrivial
(different from zero) for any $\gamma<-1$. By virtue of Lemmas 2 and
4 the restrictions of the functionals \gl{fun96} on the standard
space ${\cal{D}}(\Bbb R)\subset CL_\gamma$, $\gamma\in\Bbb R$ are
equal to zero. Hence, the supports of these functionals for any
$\gamma\in\Bbb R$ do not contain any finite real number. On the
other hand, one can represent any test function $\varphi(x)\in
CL_\gamma$, $\gamma\in\Bbb R$ for any $R>0$ as a sum of two
functions from $CL_\gamma$ in the form
\be\varphi(x)=\eta(|x|-R)\varphi(x)+[1-\eta(|x|-R)]\varphi(x),\qquad
R>0,\la{repr98}\ee where $\eta(x)\in C^\infty_{\Bbb R}$,
$\eta(x)\equiv1$ for any $x<0$, $\eta(x)\in[0,1]$ for any
$x\in[0,1]$ and $\eta(x)\equiv0$ for any $x>1$. In view of Lemmas 2
and 4 the values of the functionals \gl{fun96} for $\varphi(x)$ are
equal to their values for the second term of \gl{repr98} for any
arbitrarily large $R>0$. Hence, the values of the functionals
\gl{fun96} for a test function depend only on the behavior of this
function in any arbitrarily small (in the topological sense)
vicinity of the infinity and are independent of values of the
function in any finite interval of real axis. In this sense the
supports of the functionals \gl{fun96} for any $\gamma<-1$ consist
of the unique element which is the infinity. At last, since (1) for
any $\varphi(x)\in CL_\gamma$ and $\gamma\in\Bbb R$ the relation
\be\mathop{{\lim}_\gamma}\limits_{R\to+\infty}\eta(|x|-R)\varphi(x)=
\varphi(x)\ee holds; (2) the restrictions of the functionals
\gl{fun96} on ${\cal D}(\Bbb R)$ are zero for any $\gamma\in\Bbb R$
and (3) the functionals \gl{fun96} are nontrivial for any
$\gamma<-1$, so the functionals \gl{fun96} for any $\gamma<-1$ are
discontinuous.

In the same way one can verify that the functional
\be\mathop{{\lim}''_\gamma}_{\varepsilon\downarrow0}
{{\sin^2[{\varepsilon\over2}(x-x')]}\over{\varepsilon(x-z)(x'-z)}},
\qquad x'\in\Bbb R,\quad {\rm{Im}}\,z\ne0,\la{fun100}\ee considered
actually in \cite{soancan06}, is trivial for any $\gamma>1$ (proof
of this fact is analogous to one of Lemma 3 from Appendix of
\cite{soancan06}) and nontrivial for any $\gamma<1$ (see (71) in
\cite{soancan06}), is discontinuous for any $\gamma<1$ and its
support for any $\gamma<1$ consist of the unique element which is
infinity\footnote{It is more natural to use for the functionals
\gl{fun100} (see \cite{soancan06}) narrower spaces of test
functions $C^\infty_{\Bbb R} \cap L_2(\Bbb R;(1+|x|)^\gamma)$,
$\gamma\in\Bbb
R$. All proofs are easily adaptable  for these spaces.}.\\

\noindent{\bf Lemma 5.} {\it Suppose that the function
$\psi_0(x)=e^{z|x|/2}$ is defined for any $x\in\Bbb R$ and some purely
imaginary $z\ne0$. Then for any $\gamma>1$ and $x'\in\Bbb R$  the
relation \be\mathop{{\lim}'_\gamma}
\limits_{\varepsilon\downarrow0}\Big[\psi_0(x)\psi_0(x')
\int\limits_0^{\varepsilon(|x|+|x'|)}{{\sin t}\over t}\,dt\Big]=0\ee takes place.}\\

\noindent{\bf Proof.} It is true that \be\psi_0(x)\psi_0(x')
\int\limits_0^{\varepsilon(|x|+|x'|)}{{\sin t}\over t}\,dt\in
L_2(\Bbb R;(1+|x|)^{-\gamma})\subset CL'_\gamma,\qquad\gamma>1.\ee
Thus, to prove the lemma it is sufficient to establish that for any
$\varphi(x)\in CL_\gamma$, $\gamma>1$, the relation
\be\lim_{\varepsilon\downarrow0}
\int\limits_{-\infty}^{+\infty}\Big[\psi_0(x)\psi_0(x')
\int\limits_0^{\varepsilon(|x|+|x'|)}{{\sin t}\over
t}\,dt\Big]\varphi(x)\,dx=0\ee is valid. By virtue of the inequality
\be\Big|\int\limits_0^r{{\sin t}\over
t}\,dt\Big|\leqslant{{Kr}\over{1+r}},\qquad r\geqslant0\ee with a
constant $K>0$ independent of $r$ and Bunyakovskii inequality we
have:
$$\Big|\int\limits_{-\infty}^{+\infty}\Big[\psi_0(x)\psi_0(x')
\int\limits_0^{\varepsilon(|x|+|x'|)}{{\sin t}\over
t}\,dt\Big]\varphi(x)\,dx\Big|^2\leqslant$$
$$\int\limits_{-\infty}^{+\infty} \Big[\int\limits_0^{\varepsilon
(|x|+|x'|)}{{\sin t}\over t}\,dt\Big]^2{{dx}\over{(1+|x|)^\gamma}}
\int\limits_{-\infty}^{+\infty}|\varphi^2(x)|(1+|x|)^\gamma\,dx\leqslant$$
$$\varepsilon^2K^2\int\limits_{-\infty}^{+\infty}
{{(|x|+|x'|)^2\,dx}\over{(1+\varepsilon(|x|+|x'|))^2(1+|x|)^{\gamma_0}}}
\int\limits_{-\infty}^{+\infty}|\varphi^2(x)|(1+|x|)^\gamma\,dx=$$
$$\varepsilon^{\gamma_0-1}K^2\int\limits_{-\infty}^{+\infty}
{{(|\xi|+\varepsilon|x'|)^2\,d\xi}\over{(1+|\xi|+\varepsilon|x'|)^2
(\varepsilon+|\xi|)^{\gamma_0}}}
\int\limits_{-\infty}^{+\infty}|\varphi^2(x)|(1+|x|)^\gamma\,dx\leqslant$$
$$4\varepsilon^{\gamma_0-1}K^2\int\limits_0^{+\infty}
{{|\xi|^2+\varepsilon^2|x'|^2}\over{(1+|\xi|)^2
(\varepsilon+|\xi|)^{\gamma_0}}}\,d\xi
\int\limits_{-\infty}^{+\infty}|\varphi^2(x)|(1+|x|)^\gamma\,dx\leqslant$$
\be4K^2\Big[\varepsilon^{\gamma_0-1}\int\limits_0^{+\infty}
{{|\xi|^{2-\gamma_0}\,d\xi}\over{(1+|\xi|)^2}}+
\varepsilon|x'|^2\int\limits_0^{+\infty}{{d\xi}\over{(1+|\xi|)^2}}\Big]
\int\limits_{-\infty}^{+\infty}|\varphi^2(x)|(1+|x|)^\gamma\,dx\to0,
\qquad\varepsilon\downarrow0,\ee
where $\gamma_0=\min\{\gamma,2\}$. Thus, Lemma 5 is proved.\\

\noindent{\bf Corollary 3.} The resolution of identity \gl{re16} for
test functions from $CL_\gamma$ with $\gamma>-1$ follows from
\gl{ree15} and Lemmas 2 and 4.\\

\noindent{\bf Corollary 4.} The resolution of identity \gl{re15} for
test functions from $CL_\gamma$ with $\gamma>1$ follows from \gl{re16} and Lemma 5.\\

\noindent{\bf Lemma 6.} {\it For any $\gamma>-1$, $x'\in\Bbb R$ and
purely imaginary $z\ne0$ the relation holds, \be\mathop{{\lim}'_\gamma}
\limits_{\varepsilon\downarrow0}
\int\limits_{iz/2-\varepsilon}^{iz/2+\varepsilon}{e^{ik(|x|+|x'|)}\over{2k+iz}}\,dk=0 .\ee}\\

\noindent{\bf Proof.} In view of Corollary 2
\be\int\limits_{iz/2-\varepsilon}^{iz/2+\varepsilon}{e^{ik(|x|+|x'|)}\over{2k+iz}}\,dk\in
L_2(\Bbb R;(1+|x|)^{-\gamma})\subset CL'_\gamma,\qquad\gamma>-1.\ee
Thus, to prove the lemma it is sufficient to establish that for any
$\varphi(x)\in CL_\gamma$, $\gamma>-1$, the relation
\be\lim_{\varepsilon\downarrow0}
\int\limits_{-\infty}^{+\infty}\Big[\int\limits_{iz/2-\varepsilon}^{iz/2+\varepsilon}
{e^{ik(|x|+|x'|)}\over{2k+iz}}\,dk\Big]\,\varphi(x)\,dx=0\ee is
valid. But with the help of Corollary 2 and
Bunyakovskii inequality the proof of its validity  is quite analogous to the proof for Lemma 4.
Lemma 6 is proved.\\

\noindent{\bf Corollary 5.} The resolutions of identity \gl{re19}
and \gl{re21} for test functions from $CL_\gamma$ with $\gamma>1$
follows from \gl{re16}, \gl{soot18}, \gl{pred39} and Lemmas 2, 4 and
6.


\begin{thebibliography}{99}
\bibitem{bender} C. M. Bender and S. Boettcher, Phys. Rev. Lett. 80,
5243 (1998);\\C. M. Bender, D.C. Brody, and H. F. Jones, Phys. Rev.
Lett. 89, 270401 (2002); 92, 119902(E) (2004);\\
C.M. Bender, Rep. Prog. Phys. 70, 947 (2007).
\bibitem{acdi} A.~A.~Andrianov, F.~Cannata, J.~P.~Dedonder and M.~V.~Ioffe,
  Int.\ J.\ Mod.\ Phys.\  A {\bf 14}, 2675 (1999) .
\bibitem{cryst} M.V. Berry, J. Phys. A 31, 3493 (1998);\\M.V. Berry and D.H.J. OТDell, J. Phys. A 31, 2093
(1998);\\
K.G. Makris, R. El-Ganainy, D.N. Christodoulides, and
Z.H. Musslimani, Phys. Rev. Lett. 100, 103904 (2008);\\
S. Klaiman, U. G\"unther, and N. Moiseyev, Phys. Rev.
Lett. 101, 080402 (2008);\\
O. Bendix, R. Fleischmann, T. Kottos, and B. Shapiro,
Phys. Rev. Lett. 103, 030402 (2009);\\
S. Longhi, Phys. Rev. Lett. 103, 123601 (2009).
\bibitem{guides} A. Guo, G.J. Salamo, D. Duchesne, R. Morandotti, M.
Volatier-Ravat, V. Aimez, G. A. Siviloglou, and D. N.
Christodoulides, Phys. Rev. Lett. 103, 093902 (2009).
\bibitem{cosm} A.~A.~Andrianov, F.~Cannata and A.~Y.~Kamenshchik,
  Int.\ J.\ Mod.\ Phys.\  D {\bf 15}, 1299 (2006);\  J.\ Phys.\ A  {\bf 39}, 9975 (2006) .
\bibitem{except} M.V. Berry, Czech. J. Phys. 54, 1039 (2004);\\ W.D. Heiss,
J. Phys. A 37, 2455 (2004);\\  A.~A.~Andrianov, F.~Cannata and A.~V.~Sokolov, Nucl.\ Phys.\  B {\bf 773}, 107 (2007);\\
A.~V.~Sokolov,
  Nucl.\ Phys.\  B {\bf 773}, 137 (2007) .
\bibitem{naim} M. A. Naimark, Trudy Moskov. Mat. Obshch. 3, 181 (1954); (English transl.: Amer.
Math. Soc. Transl. (2), 16, 103 (1960)) .
\bibitem{schw} J. Schwartz, Comm. Pure Appl. Math. 13, 609 (1960).
\bibitem{pavl}B. S. Pavlov, Dokl. Akad. Nauk SSSR 146, 1267 (1962) (in Russian);\  Probl. Math. Phys. No. 1, Spectral Theory and Wave Processes, Izdat.
Leningrad Univ., 102Ц132 (1967) (in Russian).
\bibitem{krein} M. G. Krein and G. K. Langer, Dokl. Akad. Nauk SSSR 152, 39 (1963) (in Russian).
\bibitem{lyanze} V. E. Lyance, Mat. Sb. 64, 521 (1964); 65, 47 (1964); (English transl.: Amer. Math.
Soc. Transl. (2), 60, 185, 227 (1967)) .
\bibitem{sams}B. F. Samsonov, J. Phys. A 38, L397 (2005).
\bibitem{mosta}A. Mostafazadeh and H. Mehri-Dehnavi, J. Phys. A 42,
125303 (2009);\\
A. Mostafazadeh, Phys. Rev. Lett. 102, 220402 (2009).
\bibitem{ahmed} Z. Ahmed, J. Phys. A: Math. Theor. 42 (2009) 472005 .
\bibitem{gusein} G. S. Guseinov, Pramana - Journal of  Physics,  73, No. 3 (2009) 587 .
\bibitem{longhi} S. Longhi, arXiv:1001.0962 [quant-ph] .
\bibitem{prud} A.P. Prudnikov, Yu.A. Brychkov, O.I. Marichev,
{\it Integrals and series. Elementary functions} (in Russian),
Nauka, Moscow, 1981.
\bibitem{bateman1} H. Bateman, A. Erd\'elyi, {\it Higher transcendental
functions}, V. 1, Mc Graw-Hill book company, NY, Toronto, London,
1953.
\bibitem{soancan06} A.V. Sokolov, A.A. Andrianov, F. Cannata, J. Phys. A: Math. Gen.
{\bf 39} (2006) 10207.
\end{thebibliography}
\end{document}